\documentclass[%
superscriptaddress,
 amsmath,amssymb,
pre,
floatfix,]{revtex4-1}

\usepackage{graphicx}
\usepackage{dcolumn}
\usepackage{bm}
\usepackage{amssymb}
\usepackage{CJK}
\usepackage{amsmath,mathtools}
\usepackage{color}
\usepackage{hyperref}
\usepackage{psfrag}
\usepackage{stmaryrd}
\usepackage{amssymb}
\usepackage{wasysym}
\usepackage[utf8]{inputenc}
\usepackage{tabularx}
\usepackage{multirow}
\usepackage{natbib}
\usepackage{lipsum}

\begin{document}

\newcommand{\corr}[1]{{#1}}   							
\newcommand{\com}[1]{\textcolor{green}{(#1)}}
\newcommand{\bnabla}{\boldsymbol{\nabla}}
\newcommand{\bomega}{\boldsymbol{\omega}}
\newcommand{\bOmega}{\boldsymbol{\Omega}}
\newcommand{\bxi}{\boldsymbol{\xi}}
\newcommand{\la}{\left<}
\newcommand{\ra}{\right>}
\newcommand{\tS}{\tilde{S}}
\newcommand{\etal}{{\it et al.}}

\title{Wave-induced vortex recoil and nonlinear refraction}

\author{Thomas Humbert}

\affiliation{Service de Physique de l'\'Etat Condens\'e, CEA,
CNRS, Universit\'e Paris-Saclay, CEA Saclay, 91191 Gif-sur-Yvette,
France}
\affiliation{Laboratoire d'Acoustique de l'Universit\'e du Maine, UMR6613 CNRS, Univ. du Maine, F-72085 Le Mans Cedex 9, France}

\author{S\'ebastien Auma\^itre}

\affiliation{Service de Physique de l'\'Etat Condens\'e, CEA,
CNRS, Universit\'e Paris-Saclay, CEA Saclay, 91191 Gif-sur-Yvette,
France}

\author{Basile Gallet}

\affiliation{Service de Physique de l'\'Etat Condens\'e, CEA,
CNRS, Universit\'e Paris-Saclay, CEA Saclay, 91191 Gif-sur-Yvette,
France}

\date{\today}


\begin{abstract}
When a vortex refracts surface waves, the momentum flux carried by the waves changes direction and the waves induce a reaction force on the vortex. We study experimentally the resulting vortex distortion. Incoming surface gravity waves impinge on a steady vortex of velocity $U_0$ driven magneto-hydrodynamically at the bottom of a fluid layer. The waves induce a shift of the vortex center in the direction transverse to wave propagation, together with a decrease in surface vorticity. We interpret these two phenomena in the framework introduced by Craik and Leibovich (1976): we identify the dimensionless Stokes drift $S=U_s/U_0$ as the relevant control parameter, $U_s$ being the Stokes drift velocity of the waves. We propose a simple vortex line model which indicates that the shift of the vortex center originates from a balance between vorticity advection by the Stokes drift and self-advection of the vortex. The decrease in surface vorticity is interpreted as a consequence of vorticity expulsion by the fast Stokes drift, which confines it at depth. This purely hydrodynamic process is analogous to the magnetohydrodynamic expulsion of magnetic field by a rapidly moving conductor through the electromagnetic skin effect. \corr{We study vorticity expulsion in the limit of fast Stokes drift and deduce that the surface vorticity decreases as $1/S$, a prediction which is compatible with the experimental data.} Such wave-induced vortex distortions have important consequences for the nonlinear regime of wave refraction: the refraction angle rapidly decreases with wave intensity.
\end{abstract}

\maketitle

\section{Introduction} 

Wave-vortex interactions take place in various physical situations and at many different scales. Refraction of acoustic waves by  a turbulent flow is used as a way to probe the vorticity field \citep{Baudet91}. In superfluid Helium, refraction of second-sound waves by quantized vortex lines provides a way to measure the vortex line density, and interactions between quantized vortices and excitations lead to mutual friction \citep{Vinen2002}.  Wave-vortex interactions are also ubiquitous in oceanography and atmospheric flows: inertial and internal gravity waves interact with slowly evolving mean-flows, while ocean surface waves interact with the near-surface currents \citep{Phillips}. Laboratory experiments exhibited the corresponding attenuation and scattering of surface waves by background turbulent flows \citep{Green,Olmez,Gutierrez} and their refraction by localized vortices \citep{Coste99,Vivanco1999, Vivanco2004}. More recently, it was realized that swell propagating in the Pacific ocean departs from great-circle trajectories as a consequence of refraction by  mesoscale vorticity \citep{Gallet2014}.

Surprisingly, experiments on the interaction of a single vortex with incoming waves focused only on the linear level: the structure of the vortex is fixed, and one wishes to determine the refracted wave field. However, if the waves get refracted, the wave momentum flux changes direction: the vortex exerts a force on the waves, and as a reaction the wave-field exerts a force on the vortex. This reaction force deserves further attention: how does it affect the vortex? And when it does affect it, what are the laws for wave refraction in this nonlinear regime?

\begin{figure}
   \centering{\includegraphics[width=9cm]{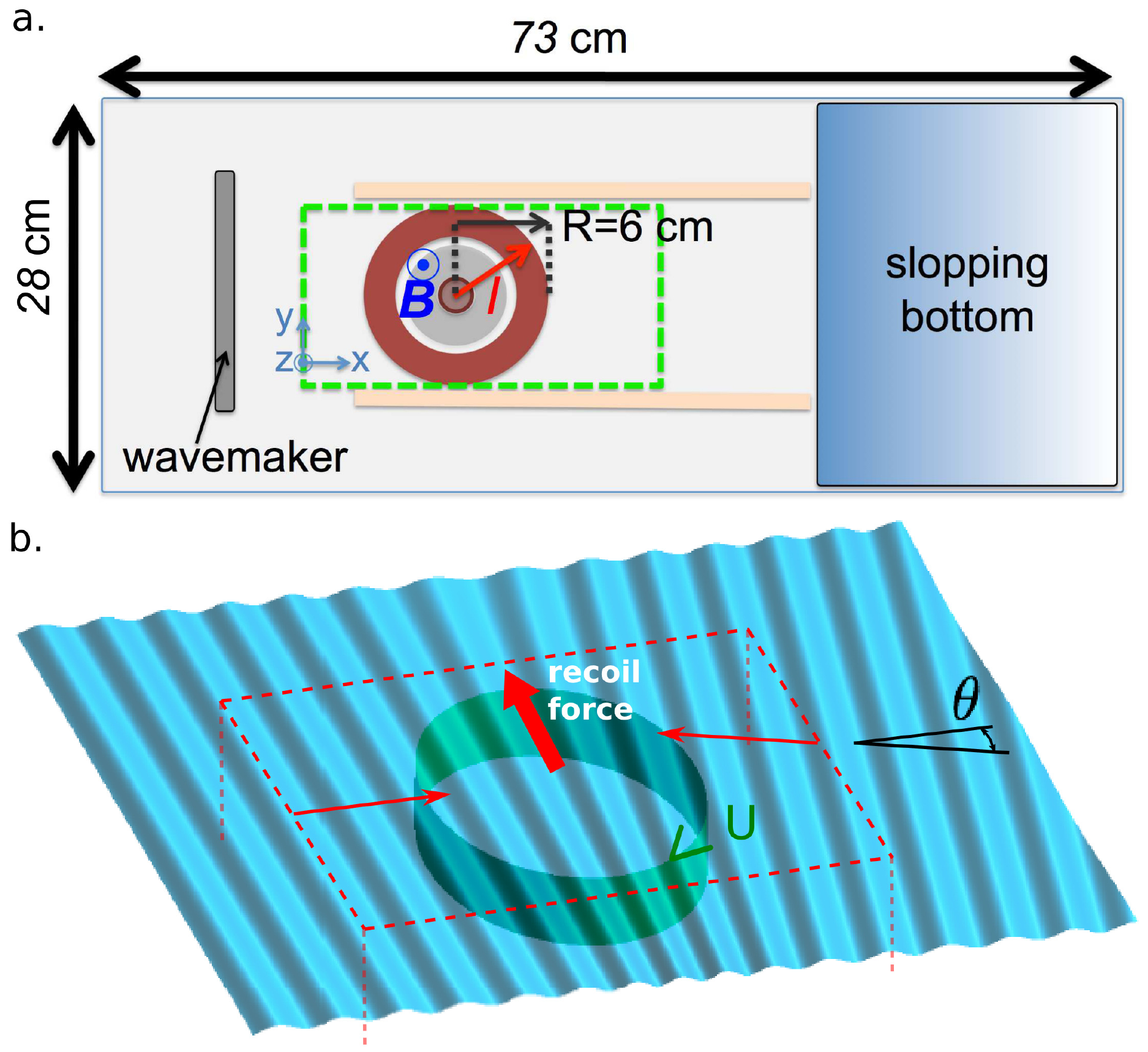}}   
    \caption{\textbf{a.} Top-view of the experimental cell, containing a height $H=3.6$ cm of copper-sulfate solution. The wavemaker is a vertically oscillating cylinder touching the free-surface. A current $I$ running between two coaxial electrodes interacts with the vertical field of a permanent magnet to produce an azimuthal Lorentz force driving the vortex. PTV in the dashed-green domain gives access to the wave-field and surface mean-flow. \textbf{b.} An azimuthal vortex (green) refracts surface waves by an angle $\theta$. Because of refraction, the radiation pressure forces due to the incoming and outgoing waves do not compensate (thin red arrows). Instead, there is a net recoil force on the vortex (thick red arrow). 
    \label{fig:schema}}
\end{figure}

Out of the above-mentioned examples, surface waves interacting with a vortex offers an opportunity to measure simultaneously the waves and the mean-flow. This constitutes the heart of the present experimental study, sketched in figure \ref{fig:schema}a: we force a steady vortex and study its distortion when subjected to incoming propagative surface gravity waves.

One can gain insight into the reaction force using the concept of radiation stresses \citep{LH1964}: surface waves carry a flux of momentum. Consider the situation sketched in figure \ref{fig:schema}b: waves impinge on a vortex that refracts them. Because of this refraction, the wave momentum flux on the dashed control surface does not vanish, resulting in a net force (see figure \ref{fig:schema}b). 
More precisely, the waves get refracted by an angle $\theta$ proportional to the typical core vorticity $\overline{\zeta}$ of the vortex \citep{Coste99bis,LLfluid,Dysthe2001,Gallet2014}.
For small $\theta$, recalling that the radiation stresses are proportional to wave intensity \citep{LH1964}, one deduces that the reaction force is quadratic in wave amplitude, proportional to the core vorticity $\overline{\zeta}$ of the mean flow and directed in the horizontal direction transverse to wave propagation, see figure \ref{fig:schema}b. 
\corr{The present study focuses on such ``direct'' recoil forces due to surface-waves overlapping with the mean-flow vorticity. When these two fields do not overlap, \citet{Buhler2003} showed that the waves nevertheless induce ``remote'' recoil forces on the vortex, due to the Eulerian return flow associated with the waves and to the weak deflection of the wave beam at second order in vortex velocity over group velocity $v_g$.}

\corr{In addition to the radiation-stress approach mentioned above, several levels of approximations have been considered to describe wave mean-flow interactions. The most general framework to include the effect of waves on the mean-flow is the Generalized Lagrangian Mean (GLM) theory \citep{Andrews,BuhlerBook}, which relies on very few assumptions and therefore allows for precise diagnostics in very general conditions. However, in their most general form the GLM equations are not closed: further approximations are needed to obtain evolution equations for the wave field that can be solved analytically or numerically \citep{CraikBook,Salmon}. Alternatively, such evolution equations for both the mean-flow and the waves can be deduced directly from the Navier-Stokes equations using standard asymptotic expansions.}

\corr{With the goal of describing Langmuir circulations, \citet{Craik1976} followed such an asymptotic route and derived a particularly simple closed equation for the evolution of flows subject to weak surface waves. Expanding the vorticity equation in powers of the weak wave-slope, they showed that the vorticity ${\bomega}$ of the mean-flow ${\bf U}$ obeys the simple equation}:
\begin{equation}
\partial_t \bomega = \bnabla \times \left[ \left( {\bf U} + {\bf u}_s \right) \times \bomega \right] + \nu \bnabla^2 \bomega  \, ,\label{CLint}
\end{equation}
where ${\bf u}_s$ denotes the Stokes drift associated to the wave field. This equation corresponds to the usual vorticity equation with an additional term proportional to ${\bf u}_s$ on the right-hand side, which arises as a consequence of the time-averaged wave-induced Reynolds stresses. This additional term corresponds to a force ${\bf u}_s \times \bomega$ on the right-hand side of the Navier-Stokes equation, up to a gradient: it is indeed quadratic in wave amplitude, proportional to the local vorticity of the mean flow, and transverse to the direction of wave propagation, given by ${\bf u}_s$. The Craik-Leibovich framework is therefore consistent with the insight gained from the concept of wave stresses. A detailed comparison of the radiation-stress and vortex-force representations of wave-averaged effects is given in \citet{Lane}. 
The vortex force therefore offers an efficient way to include the influence of surface waves in the mean-flow vorticity equation. We stress the rather broad range of parameters covered by this approach: first, the Craik-Leibovich equation (\ref{CLint}) does not require any scale separation between the wavelength $\lambda$ and the scale $R$ of the mean-flow, in contrast with ray-tracing approaches. Second, the original assumption of \citet{Craik1976} that the mean flow arises at second order in wave-slope stems mostly from asymptotic elegance, so that all the terms of equation (\ref{CLint}) arise at the same order in a perturbative expansion. \corr{Subsequently, Leibovich realized that the Craik-Leibovich equation can be derived as a limiting situation of the more general GLM equations \cite{Leibovich80}. Craik then took advantage of the GLM equations to extend the approach to fast unidirectional mean flows, with velocities of the order of the phase speed of the waves \cite{Craik1982}. As detailed by Phillips, such combinations of surface waves and fast unidirectional flows also lead to the formation of Langmuir circulations \cite{Phillips94,Phillips05}. When the mean-flow is fully three-dimensional and has a magnitude comparable to the wave orbital velocity -- slow as compared to the phase speed but fast as compared to the standard Craik-Leibovich scaling -- we show in appendix \ref{appendix} that equation (\ref{CLint}) still applies: the vortex-force remains the first wave-induced correction to the usual vorticity equation, but it is a subdominant term if the mean-flow is too fast.}

Apart from Langmuir circulations, for which it was designed, the Craik-Leibovich equation (\ref{CLint}) provides a simple way to study theoretically the influence of surface waves on background turbulence \citep{Teixeira2002} as well as on the global oceanic circulation \citep{McWilliams1999}. In the following we use it extensively to interpret the experimental data.

\section{Experimental setup} 
The experimental setup is sketched in figure \ref{fig:schema}a: a cell of length 73 cm and width 28 cm contains a layer of copper sulfate solution of depth $H=3.6$ cm. On one side, an electromagnetic shaker drives a wavemaker touching the free-surface. The vertical sinusoidal motion of the wavemaker at constant frequency $f=1/T$ generates surface waves of amplitude $a$ up to 6 mm that propagate across the cell. A sloping bottom at the other side of the cell ensures that the incoming waves are damped without reflection. In the central region of the cell, two vertical walls channel the waves while allowing for recirculating flows close to the side walls.

\begin{figure*}
    \centerline{\includegraphics[width=15cm]{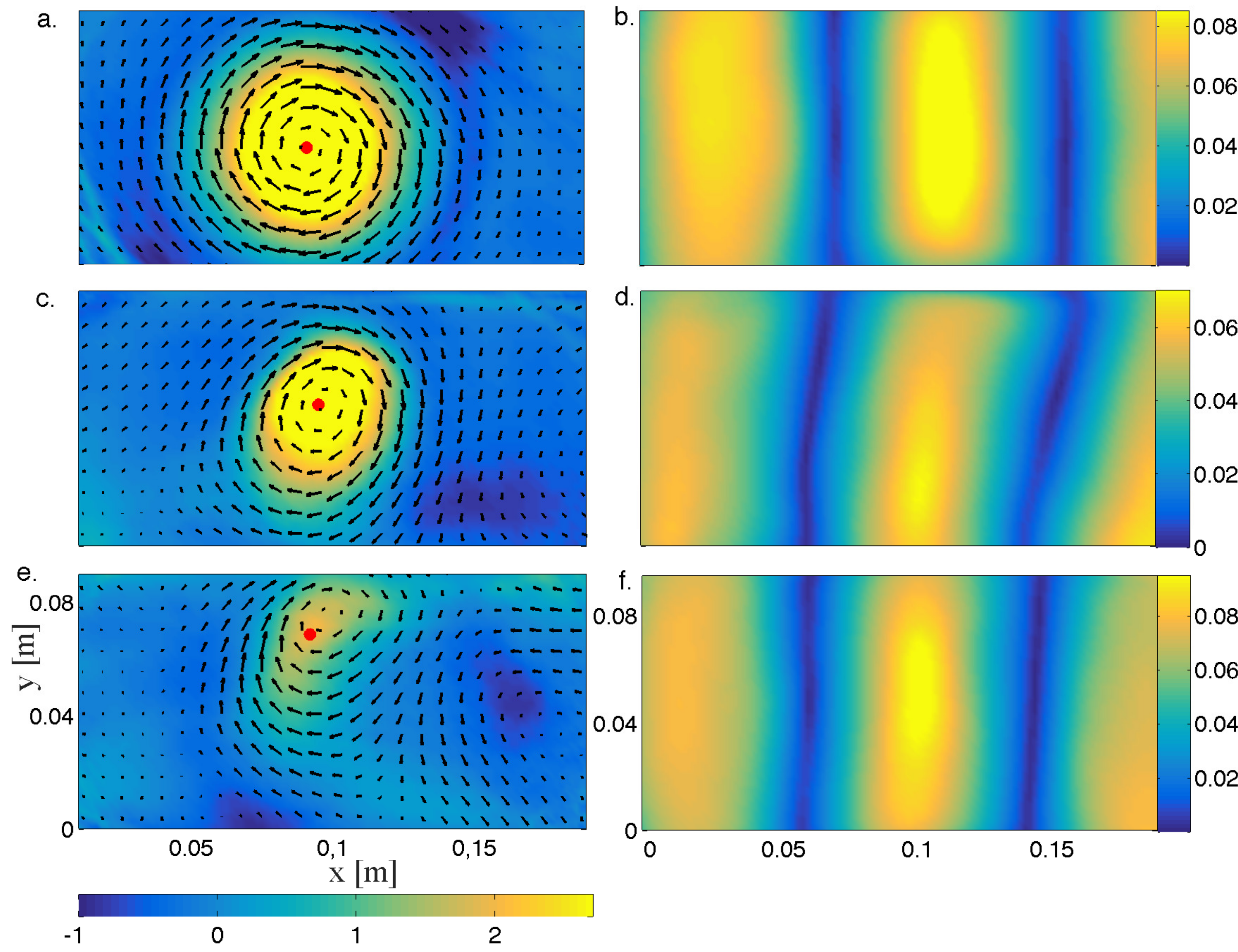}}
    \caption{\textbf{Left:} snapshots of the surface mean-flow ${\bf U}$ for $U_0=66$ mm.s$^{-1}$. The arrows are the velocity field in arbitrary units, and color codes for the negative of vertical vorticity, $-\zeta$, in inverse seconds. \textbf{Right:} instantaneous wave field. Color codes for $|\tilde{\bf u}|^2$ in m$^2$.s$^{-2}$. The different panels correspond to: a, vortex only; b, waves only; c and d, moderate waves lead to a small transverse displacement of the vortex center together with strong wave refraction ($U_s=7$ mm.s$^{-1}$); e and f, energetic waves induce a large vortex displacement together with reduced wave refraction ($U_s=11$ mm.s$^{-1}$). The axes in panels e and f apply to the entire column. The colorbar of panel e applies to panels a, c and e.
    \label{fig:snapshots}}
\end{figure*}

The vortical flow is generated magneto-hydrodynamically: two coaxial copper electrodes and a permanent magnet are flush with the bottom of the cell. The magnet produces a vertical magnetic field $B_0 \simeq 0.3$ T in its vicinity. We impose an electrical current $I$ between the two electrodes: the combination of the radial current with the vertical magnetic field induces an azimuthal Lorentz force that drives a vortex. Between the two electrodes, a small insulating cylinder (inner diameter $=2.7$ cm, outer diameter $=6.2$ cm, height $=1.3$ cm) anchors the base of the vortex. In this configuration, one observes an approximately steady and mostly azimuthal flow, with weak poloidal recirculation (see figure \ref{fig:snapshots}a). We define the typical velocity of the vortex as:
\begin{equation}
U_0=\frac{1}{R} \sqrt{\iint_{x,y} |{\bf U}|^2 \mathrm{d}x \mathrm{d}y} \, ,\label{defU}
\end{equation}
where ${\bf U}$ denotes the surface velocity field and the integral is over the fluid surface. The shape of the vortex is roughly independent of $I$ and its typical velocity $U_0$ evolves as $\sqrt{I}$. This behavior is typical of flows driven by a steady body-force and corresponds to the high-Reynolds-number or ``turbulent'' scaling regime where the nonlinear term of the Navier-Stokes equation balances the driving force \citep{Borue}. In the present magnetohydrodynamic (MHD) situation, balancing the nonlinear term with the Lorentz force yields the following scaling-law \citep{Boisson}:
\begin{equation}
U_0 = {\cal C} \sqrt{\frac{I B_0}{R \rho}} \, . \label{ODGU}
\end{equation}
Here $R=6$ cm is the outer radius of the outer electrode, which we use as the typical length scale, $\rho$ is the fluid density and ${\cal C}$ is a dimensionless constant. 
The experimentally measured prefactor ${\cal C}=1.3$ is close to unity, which confirms that the Lorentz force and nonlinear terms of the Navier-Stokes equation are indeed of similar magnitude.



The surface flow velocity is accessed by tracking particles floating on the free-surface. The particle-tracking velocimetry (PTV) system is phase-locked with the oscillation of the wavemaker, and records eight velocity fields per wave period: coherent-averaging leads to maps of the flow at eight different phases during the oscillation period. The velocity field of the floating particles can be decomposed into several contributions: the mean Eulerian flow is denoted as ${\bf U}(x,y,z=0)$; the wave Eulerian flow $\tilde{\bf u}(x,y,t)$ is periodic in time and has mean zero. Because the waves have a finite amplitude, particles on the free-surface are subject to Stokes drift: the mean zero Eulerian velocity $\tilde{\bf u}(x,y,t)$ induces a non-zero Lagrangian drift along $x$ of the surface particles, of magnitude $U_s=2 \pi a^2 f k$, where $k=2\pi / \lambda$ is the wavenumber.
\corr{Particle-based velocimetry is usually of two kinds: one can follow a given particle over a long time to obtain the trajectory of the particle. Such particle trajectories are subject to the full Stokes drift and provide accurate measurements of the Lagrangian velocity. By contrast, one can focus on a fixed Eulerian position $(x,y,z)$ and study at each time the velocity of the particular particle that passes at this location: this leads to the Eulerian velocity field.} In this respect, the surface-particle-tracking system is a mixed Eulerian-Lagrangian measurement: it gives access to the velocity of the particle passing at a given (Eulerian) horizontal position $x$ and $y$, but at a (Lagrangian) height that varies sinusoidally in time because of the oscillation of the free surface. In this configuration, denoting the Lagrangian displacement as $\bxi=(\xi_x, \xi_y, \xi_z)$, with $\tilde{\bf u}=\partial_t \bxi$, we do not sense the full Stokes drift \corr{$U_s {\bf e}_x = (\bxi \cdot \bnabla) \, \tilde{\bf u}$, but only $\xi_z \partial_z \tilde{u} \, {\bf e}_x$, where $\tilde{u}=\tilde{\bf u}\cdot {\bf e}_x$}. For monochromatic deep-water surface gravity waves, the latter term equals $\frac{U_s}{2} {\bf e}_x + \tilde{\bf u}_2(x,y,t)$, where $\tilde{\bf u}_2$ is a second-harmonic term: it has mean zero and oscillates at frequency $2f$. The velocity ${\bf u}$ of the floating particles is finally:
\begin{equation}
{\bf u} (x,y,t) = {\bf U}(x,y,z=0) + \tilde{\bf u}(x,y,t) + \frac{U_s}{2} {\bf e}_x + \tilde{\bf u}_2(x,y,t)\, .
\end{equation}
Averaging this velocity field over all eight measured phases and subtracting the Stokes drift term gives the mean surface flow ${\bf U}(x,y,z=0)$. 
In the following, we use the same notation ${\bf U}$ to denote both this mean-flow measured at the surface and the full 3D mean-flow appearing in the equations, e.g. in (\ref{CLint}).
Subtracting the velocity fields measured at phases of opposite polarity and dividing by two leads to the wave field: $\tilde{\bf u}(x,y,t) =  [{\bf u} (x,y,t)-{\bf u} (x,y,t+T/2)]/2$. Surface PTV therefore gives access to both the wave-field and the Eulerian mean flow with the same measuring technique.

When studying wave mean-flow interactions, a key issue is to ensure that the wave-generation system does not induce significant Eulerian streaming, which would directly affect the vortex.  We therefore restrict attention to situations where the Eulerian streaming flow (measured when only the waves are on) is below $10\%$ of the vortex velocity (measured when only the vortex is on). This criterion is quite demanding and led us to fix the wave frequency to $f=2.8$ Hz, which corresponds a wavelength $\lambda=16$ cm in the absence of mean-flow.


\section{Vortex distortion}

For each set of parameters, we first measure the vortex without waves (see figure \ref{fig:snapshots}a) and determine its typical velocity $U_0$ according to (\ref{defU}), where the integral is performed over the entire image. $U_0$ also corresponds to the typical velocity at the bottom of the tank, regardless of the wave amplitude, and it turns out to be a slightly more accurate control parameter than the current $I$. We then measure the waves only (without vortex) and determine their amplitude and associated Stokes drift ${\bf u}_s$, before checking that Eulerian streaming is negligible. We then turn on both the vortex and the waves, and we measure the modified mean-flow ${\bf U}$ and wave field $\tilde{\bf u}$.
Figure \ref{fig:snapshots} shows the vertical vorticity $\zeta=(\bnabla \times {\bf U})\cdot {\bf e}_z$ of the mean-flow for increasing wave intensity: the reaction force sketched in figure \ref{fig:schema}b induces a distortion of the vortex core together with a shift of the vortex center in the direction transverse to wave propagation. A more surprising observation is that the vorticity decreases dramatically for increasing wave intensity. Similar observations are made if the current is decreased while the wave amplitude is held constant.

\begin{figure}
\begin{center}
  {\includegraphics[width=8cm]{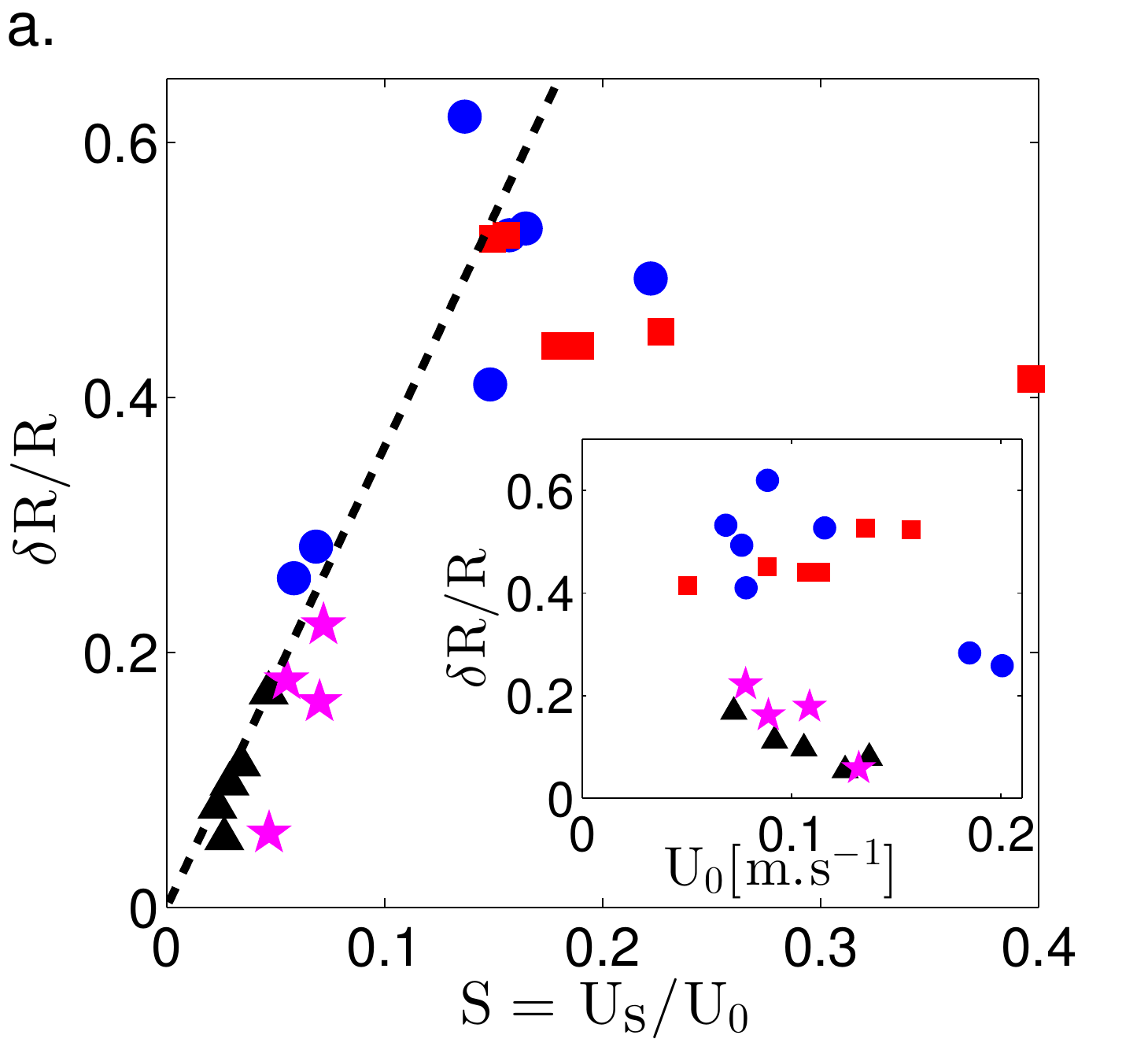}}
  {\includegraphics[width=8cm]{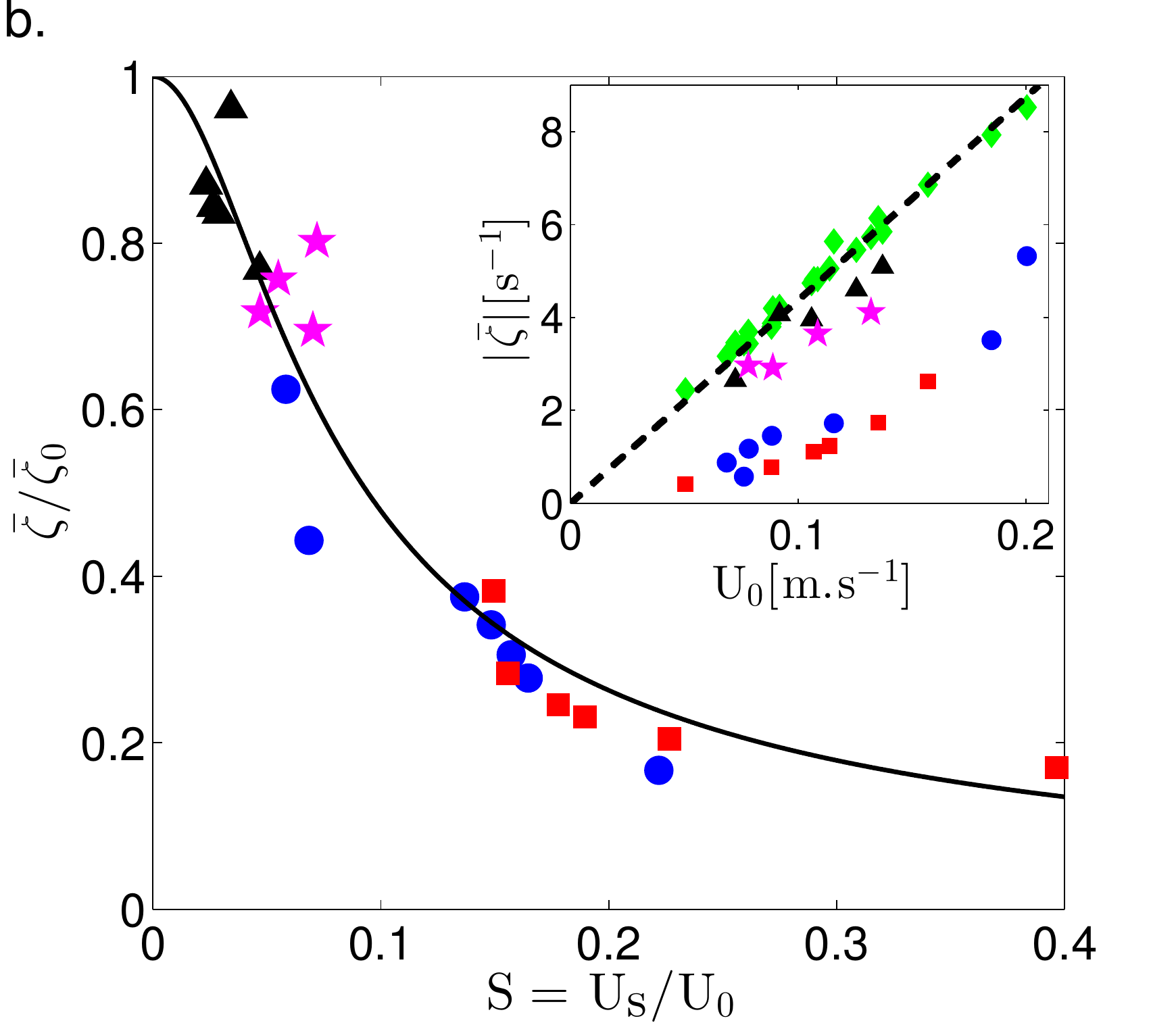}}
  \end{center}
    \caption{ {\bf a.} Relative shift of the vortex center position, plotted as a function of the base-flow velocity $U_0$ measured without waves (inset) and as a function of the rescaled Stokes drift $S$ (main figure). The dashed-line is the linear scaling-law predicted by the vortex-line model. {\bf b.} Inset: Core vorticity $\overline{\zeta}$ as a function of the control velocity $U_0$. The dashed-line is an eye-guide. The main figure shows $\overline{\zeta}$ rescaled by the core vorticity without waves $\overline{\zeta}_0$, as a function of the dimensionless Stokes drift. \corr{The solid line is the prediction (\ref{defhatzeta}) of the simplified expulsion model for the typical surface vorticity $\hat{\zeta}$ normalized by its value for $S=0$.} Symbols are: $\diamond$, $U_s \leq 0.16$ mm.s$^{-1}$ ; $\triangle$, $3.1 \leq U_s \leq 3.4$ mm.s$^{-1}$ ; $\star$, $5.6 \leq U_s \leq 7$ mm.s$^{-1}$ ; $\bullet$, $11 \leq U_s \leq 18$ mm.s$^{-1}$; $\square$, $20 \leq U_s \leq 24$ mm.s$^{-1}$. \label{fig:scalings}}
\end{figure}

To make these observations quantitative, we estimate the vortex center as the centroid of the region where $\zeta$ exceeds its core rms value (red dots in figure \ref{fig:snapshots}). We then plot in the inset of figure \ref{fig:scalings}a the departure $\delta R$ of the vortex center position, as compared to its position in the absence of waves. For fixed control velocity $U_0$ (fixed current $I$), $\delta R$ increases with wave intensity. The typical core vorticity is defined as:
\begin{equation}
\overline{\zeta}=\frac{1}{R} \sqrt{\iint_{x,y} \zeta^2 \mathrm{d}x \mathrm{d}y} \, ,
\end{equation}
where the integration is over the entire vortex core, i.e., over the central region of negative vorticity visible in snapshots \ref{fig:snapshots}a, c, e.
In the inset of figure \ref{fig:scalings}b, we show
 $\overline{\zeta}$ as a function of the control velocity $U_0$. For a given $U_0$, $\overline{\zeta}$ rapidly decreases with wave intensity.\\

\section{Interpretation \label{sec:interpretation}}

We now wish to explain the two wave-induced phenomena reported above: the shift of the vortex center and the decrease in surface vorticity. The starting point is the Craik-Leibovich equation (\ref{CLint}) for the mean-flow vorticity ${\bomega}= \bnabla \times {\bf U}$:
\begin{equation}
\partial_t \bomega = \bnabla \times \left[ \left( {\bf U} + U_s e^{2kz}{\bf e}_x \right) \times \bomega \right] + \nu \bnabla^2 \bomega + \bnabla \times {\bf F} \, , \label{CLeq}
\end{equation}
where the Lorentz force per unit mass ${\bf F}$ is proportional to $I$ and we have substituted the expression for the Stokes drift of a monochromatic deep-water wave, ${\bf u}_s =U_s e^{2kz} {\bf e}_x$. Although this equation was originally derived for mean-flows much slower than the wave orbital velocity, we show in appendix \ref{appendix} that it remains valid when the two are comparable, as is the case in our experiment. 

\corr{Because of the rapid spatial decay of the magnetic field and electrical currents, the Lorentz force ${\bf F}$ is confined near the bottom of the tank. The balance between this force and the nonlinear term sets the velocity and vertical vorticity in this bottom region. Alternatively, this balance can be modeled though the use of a turbulent viscosity $\nu_t$ proportional to the typical scale and velocity of the fast turbulent vortex motion imposed by the forcing near the bottom of the tank:} 
\begin{equation}
\nu_t = c \, R U_0 \, , \label{defnut}
\end{equation}
where $c$ is a dimensionless prefactor. The turbulent viscous term then compensates the Lorentz force to yield the scaling-law (\ref{ODGU}). Away from the bottom forcing region, the vorticity evolves according to the unforced Craik-Leibovich equation (\ref{CLeq}) with  ${\bf F}={\bf 0}$. In steady state, the vortex force is balanced by the combination of the nonlinear term and the turbulent viscous one. In the following we study independently the effect of these two terms.

\subsection{Self-induction and vortex displacement \label{self}}

\begin{figure}
\begin{center}
 {\includegraphics[width=8.5cm]{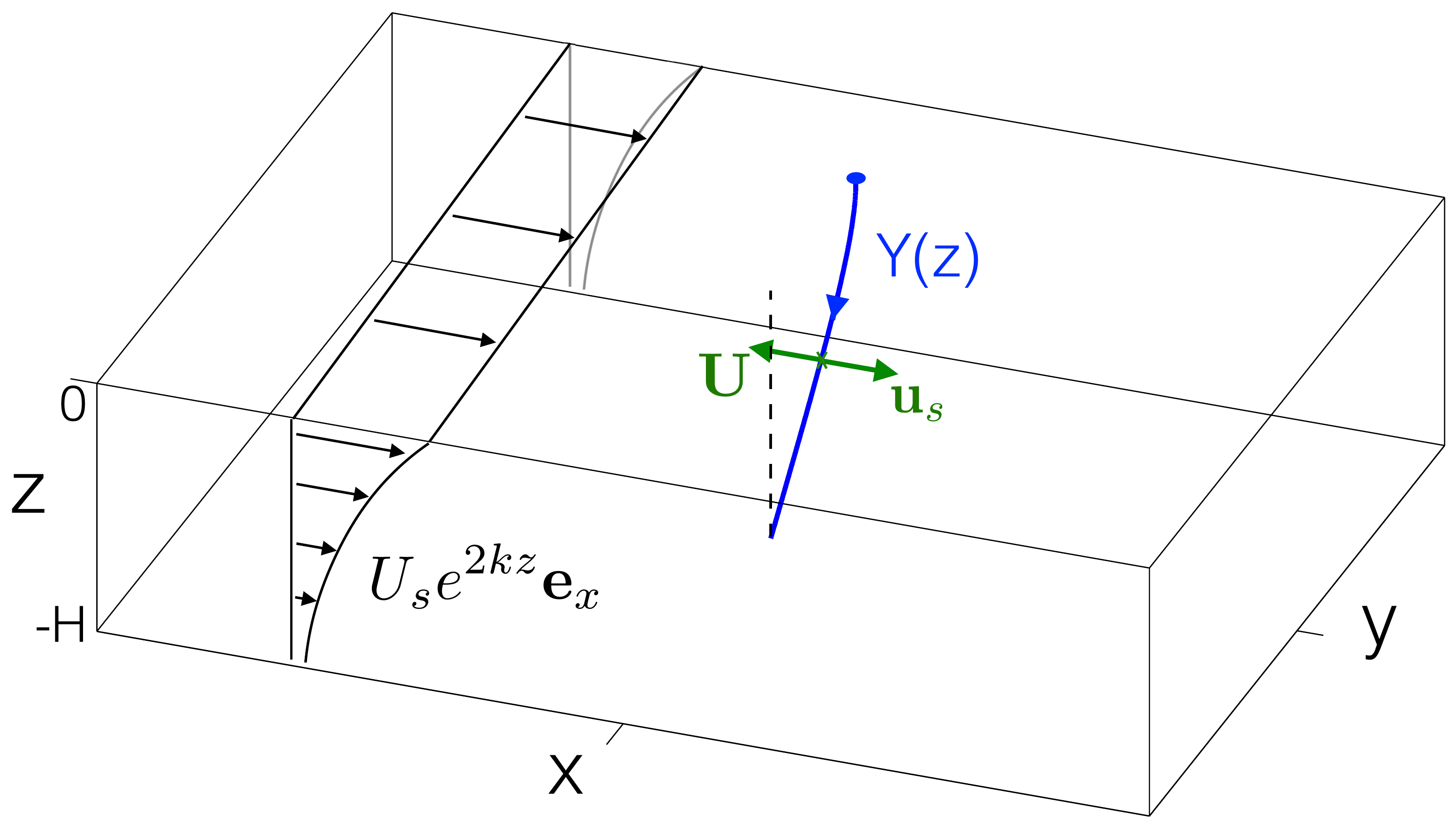}}
  {\includegraphics[width=6.5cm]{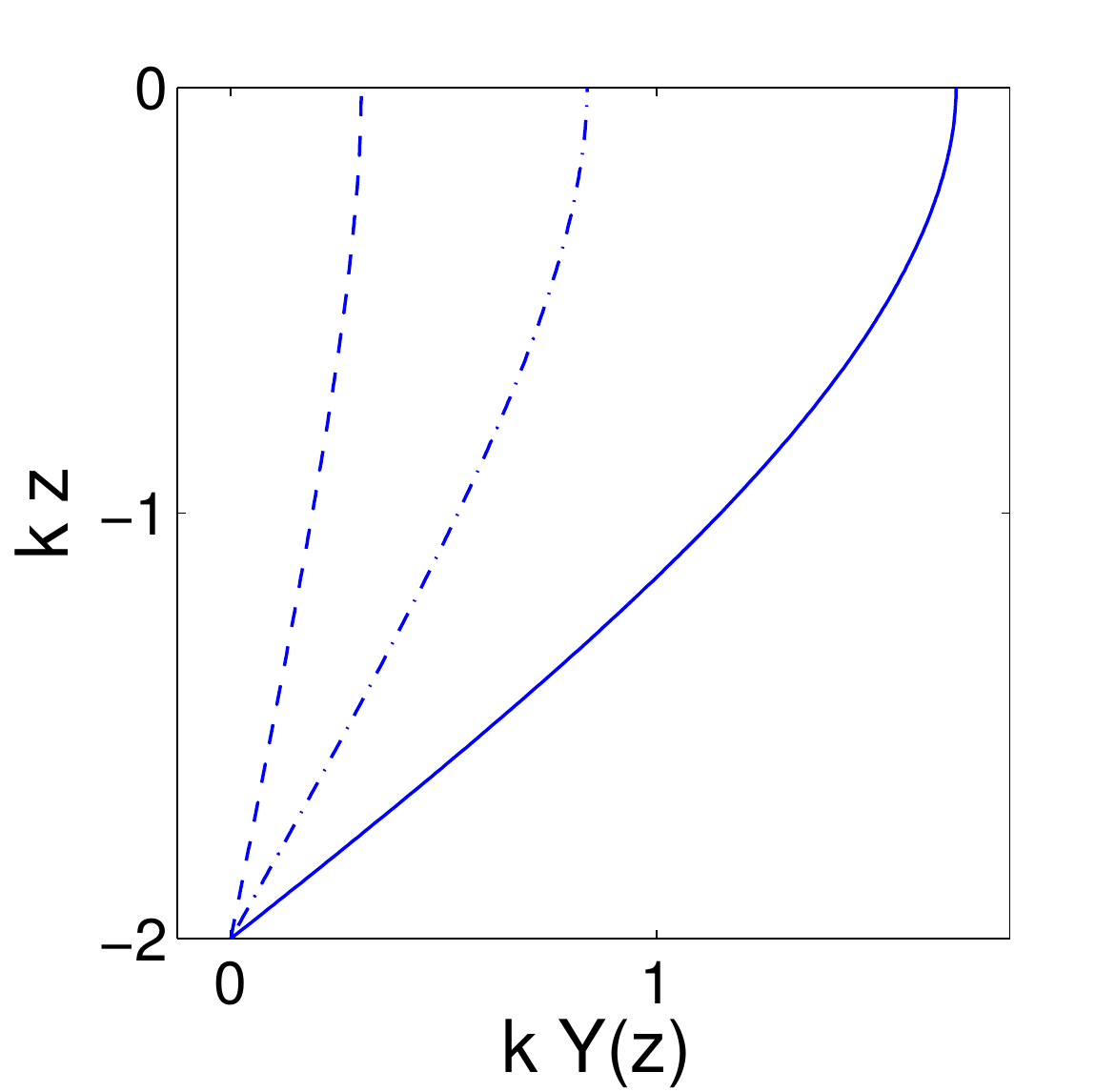}}
 \end{center}
    \caption{{\bf Left:} surface gravity waves impinging on a vortex line induce a displacement of the line transverse to the direction of wave propagation. The equilibrium position of the line is determined by the balance between the local Stokes drift ${\bf u}_s$ and the self-induced velocity ${\bf U}$ of the line. {\bf Right:} equilibrium shape (\ref{lineshape}) of the vortex line, plotted in dimensionless form for $kH=2$ and several values of $\tS = U_s/(2 k \Gamma)$: dashed line, $\tS=0.2$; dash-dotted line, $\tS=0.5$; solid line, $\tS=0.8$. \label{fig:line}}
\end{figure}

Let us focus on the first consequence of waves impinging on a vortex: the surface displacement of the vortex center. This transverse shift of the vortex results from the interplay between Stokes drift and nonlinear advection. To illustrate this phenomenon, we design a minimal model focusing on these two ingredients. Consider a vortex line in an inviscid fluid occupying $z \in [-H,0]$, as sketched in figure \ref{fig:line}. Waves of wavenumber $k$ propagate along $x$ on top of the fluid layer, producing a Stokes drift $U_s e^{2kz} {\bf e}_x$. The vortex line is a simplified model for a thin vortex tube of core radius $a$ and circulation $\Gamma$, and when the typical velocity $\Gamma/a$ is less than the wave orbital velocity, the Craik-Leibovich framework is applicable (see appendix \ref{appendix}). In this framework, the vortex line is advected by the Stokes drift as well as by the self-induced velocity arising from the curvature of the line. In steady state, the two balance at each point of the line, i.e., there is an equilibrium between vortex force and nonlinear advection. The Biot-Savart law indicates that the self-induced velocity is perpendicular to the plane of the line. In steady state, this self-induced velocity is directed along $x$ to compensate the Stokes drift, and the line therefore lies in the $(Oyz)$ plane. This is a first indication that the (inviscid) balance between vortex force and nonlinear advection leads to a transverse displacement of the vortex center (i.e., along $y$) \cite{Schwarz85}. We therefore parametrize the shape of the vortex line as $Y(z)$, as sketched in figure \ref{fig:line}. To obtain an analytically tractable model, we assume that the core radius $a$ is much less than the radius of curvature of the vortex line and consider the Local Induction Approximation \citep{Arms,Saffman}: the self-induced velocity is directly proportional to the local curvature $\kappa$ of the vortex line, up to logarithmic prefactors that we neglect in this simple model. The balance between Stokes drift and self-induced velocity therefore reads:
\begin{equation}
\Gamma \kappa \, {\bf e}_x = U_s e^{2 k z} {\bf e}_x\, ,
\end{equation}
and substituting the expression of the curvature $\kappa$ yields:
\begin{equation}
\Gamma \frac{Y''(z)}{[1+Y'^2(z)]^{3/2}}= \pm U_s e^{2 k z} \, . \label{tempcurv}
\end{equation}

The sign depends on the direction of the vortex line. We proceed with a positive sign, which corresponds to the situation of figure \ref{fig:snapshots} where vorticity points towards negative $z$ initially. The vortex line originates from $(x,y)=(0,0)$ at the bottom of the fluid domain, which models the top of the forcing region, and it connects perpendicularly to the top free surface. The boundary conditions are therefore $Y(-H)=0$ and $Y'(0)=0$.
(\ref{tempcurv}) is a separable equation for the variables $Y'$ and $z$, and after two integrations we obtain the equilibrium shape of the vortex line:

\begin{eqnarray}
 Y(z)  &  =  & \frac{1}{2k} \Bigg\{ \arcsin \left[ \tS (e^{-2kH}-1 ) \right]   \label{lineshape}  - \arcsin \left[ \tS (e^{2kz}-1 ) \right] + \left.  \frac{\tS}{\sqrt{1-\tS^2}}  \Bigg[ 2k(z+H) \right.  \\
\nonumber   &  & - \ln   \left.  \left.   \frac{1+ \tS^2 (e^{2kz}-1 ) + \sqrt{1-\tS^2} \sqrt{1-\tS^2(e^{2kz}-1 )^2}}{1+ \tS^2 (e^{-2kH}-1 ) + \sqrt{1-\tS^2} \sqrt{1-\tS^2(e^{-2kH}-1 )^2}}   \right]  \right\} \, ,
\end{eqnarray}


where $\tS = U_s/(2 k \Gamma)$. We plot this solution in figure \ref{fig:line} for several intensities of the Stokes drift. \corr{Although the Stokes drift is rather confined near the fluid surface, one can notice that the vortex line is affected over the entire fluid depth.} The displacement of the vortex center at the free-surface is given by (\ref{lineshape}) evaluated at $z=0$. When the Stokes drift is small, $\tS \ll 1$, Taylor expansion shows that this displacement is proportional to the dimensionless Stokes drift $S=U_s/U_0$, up to geometrical prefactors. We confront this prediction to the experimental data in figure \ref{fig:scalings}a: plotting the relative displacement $\delta R /R$ against $S$ leads to a good collapse of the data. As predicted from the simple vortex line model, the vortex displacement increases linearly in $S$ for small $S$. Because of the finite size of the wave channel this increase then saturates for larger $S$. This simple model is therefore in qualitative agreement with the experimental data, even though the shallow experimental vortex strongly differs from a thin vortex line: for small $S$, the vortex center is shifted in the direction perpendicular to wave propagation by a distance proportional to $S$. 

\subsection{Skin effect and the expulsion of vorticity \label{skin}}


\corr{We now turn to the second consequence of waves impinging on a vortex: the reduction of surface vorticity. The mechanism at play} is that fast Stokes drift prevents the bottom vertical vorticity from reaching the fluid surface, in a fashion similar to the skin effect inside electrically conducting media \citep{Jackson}. Indeed, away from the forcing region, equation (\ref{CLeq}) is identical to the induction equation for the evolution of the magnetic field inside an electrically conducting fluid: $\bomega$ plays the role of the magnetic field, $\nu$ plays the role of magnetic diffusivity and ${\bf U} + U_s e^{2kz}{\bf e}_x$ represents the total velocity of the conducting fluid \citep{Batchelor}. When this velocity is fast, the magnetic field is expelled from the rapidly moving fluid and remains confined to a skin layer at depth \citep{Moffatt}. In a similar fashion, equation ({\ref{CLeq}}) indicates that fast Stokes drift expels vorticity from the top region of the fluid layer and confines it to the bottom part of the tank. 

\corr{To further illustrate this expulsion mechanism in a simple configuration, we make further simplifying assumptions. First, we focus on the steady-state solution and we consider a shallow layer of fluid, so as to neglect horizontal derivatives as compared to vertical ones. Second, we replace the molecular viscosity by the turbulent one, $\nu_t$, given by (\ref{defnut}). 
Finally and most importantly, we discard the nonlinear term of equation (\ref{CLeq}), keeping only the vortex force and the turbulent viscous one. Strictly speaking, this crude approximation holds only wherever ${\bf u}_s \gg {\bf U}$. Although this inequality is not satisfied experimentally, focusing on this limiting situation gives some useful insight into the fundamental mechanism of vorticity expulsion by the Stokes drift.}

Under these assumptions, the vertical component of the Craik-Leibovich vorticity equation reduces to:
\begin{equation}
U_s e^{2kz} \partial_x \zeta = \nu_t \partial_{zz} \zeta \, . \label{eqzeta}
\end{equation}
At the bottom of the domain the velocity and vorticity fields must match those of the forcing region. We model this imposed bottom vorticity by a simple Gaussian profile:
\begin{equation}
\zeta(x,y,z=-H)=\zeta_0 \exp\left({-4 \, \frac{x^2+y^2}{R^2}}\right) \, . \label{CLbottom}
\end{equation}

Equation (\ref{eqzeta}) is an advection-diffusion equation for the vertical vorticity $\zeta$, which behaves as a passive tracer: the tracer concentration is imposed by the boundary condition (\ref{CLbottom}), and it diffuses from the bottom of the cell upwards while being advected by the Stokes drift. This leads to a shift of the vortex center {\it along} the direction of wave propagation, together with a spreading of the surface vorticity. This qualitative difference between the model and the experiment originates from the nonlinear term, which is responsible for the experimental vortex being shifted in the direction transverse to wave propagation. Neglecting this nonlinear term is justified only in the limit ${\bf u}_s \gg {\bf U}$ for a shallow fluid layer: we checked through direct numerical simulation of equation (\ref{CLeq}) that, in this regime, the shift of the vortex center is indeed along the direction of wave propagation (not shown).

For a screened vortex imposed at the bottom of the fluid layer, regions of positive and negative vertical vorticity spread and overlap at the surface, which results in important cancellations: the vorticity is thus expelled from the surface. By contrast, for the simple Gaussian model (\ref{CLbottom}), the vorticity is single-signed: the central spot of vorticity spreads at the surface, which also results in reduced maximum vorticity at the surface.

In order to study the displacement and spreading of the surface vorticity, we consider the first moments in $x$ of the vorticity distribution:
\begin{eqnarray}
M_0(z) & = & \iint \zeta(x,y,z) \mathrm{d}x \mathrm{d}y \, , \label{defM0}\\
M_1(z) & = & \iint x \zeta(x,y,z) \mathrm{d}x \mathrm{d}y \, , \\
M_2(z) & = & \iint x^2 \zeta(x,y,z) \mathrm{d}x \mathrm{d}y \, , \label{defM2}
\end{eqnarray}
where the integrals in the horizontal directions extend over the entire domain. The values of these moments are imposed at the bottom of the domain by the boundary condition (\ref{CLbottom}), while their first $z$-derivative vanishes at the free surface. Multiplying equation (\ref{eqzeta}) respectively by $x^0$, $x^1$ and $x^2$ before integrating over the horizontal coordinates leads to a hierarchy of equations that allows to successively compute the various moments:
 \begin{eqnarray}
M_0''(z) & = & 0 \, , \label{eqM0}\\
M_1''(z) & = & - \frac{U_s e^{2kz}}{\nu_t} M_0(z) \, , \\
M_2''(z) & = & - \frac{2 U_s e^{2kz}}{\nu_t} M_1(z) \, , \label{eqM2}
\end{eqnarray}
where the prime denotes differentiation with respect to $z$. The dimensionless combination governing the vorticity spreading is $U_s R / \nu_t$, which is proportional to $S=U_s/U_0$ after substituting (\ref{defnut}). From the hierarchy of equations (\ref{eqM0}-\ref{eqM2}), one can readily see that $M_0$ is independent of $S$, $M_1 (z)$ is proportional to $S$ and $M_2(z)$ departs from its bottom value $M_2(-H)$ as $S^2$.

We introduce the centroid $X$ of the surface vorticity distribution and its typical width $\delta X$ as:
 \begin{eqnarray}
X & = & \frac{\iint x \zeta(x,y,0) \mathrm{d}x \mathrm{d}y }{\iint \zeta(x,y,0) \mathrm{d}x \mathrm{d}y } = \frac{M_1(0)}{M_0(0)} \, ,  \\
(\delta X)^2 & = & \frac{\iint (x-X)^2 \, \zeta(x,y,0) \mathrm{d}x \mathrm{d}y }{\iint \zeta(x,y,0) \mathrm{d}x \mathrm{d}y } \\
\nonumber & = & \frac{M_2(0)}{M_0(0)} - X^2 \, ,
\end{eqnarray}
before defining the typical surface vorticity as:
 \begin{eqnarray}
\hat{\zeta} = \zeta_0 \frac{R}{\delta X} \, .\label{defhatzeta}
\end{eqnarray}
One can easily check that, for large $S$, $\delta X$ is proportional to $S$ and therefore $\hat{\zeta}$ scales as $S^{-1}$. The expressions of the moments $M_0$, $M_1(z)$, $M_2(z)$ needed to compute $X$, $\delta X$ and $\hat{\zeta}$ are given in appendix \ref{appmoments}.

In the experiment, the typical surface vorticity scales like $\overline{\zeta}$. In figure \ref{fig:scalings}b, we therefore plot the rescaled surface vorticity $\overline{\zeta}/\overline{\zeta}_0$ as function of the dimensionless Stokes drift $S$. This representation leads to a collapse of the experimental data onto a single master curve, which indicates that $S$ is indeed the relevant control parameter. It also confirms that molecular viscosity is irrelevant in this high-Reynods-number regime, since there is no specific dependence on Reynolds number. We can compare this master curve to the prediction $\hat{\zeta}$ of the simplified expulsion model: we use the experimental values $kR=2.3$ and $H/R=0.6$, and we fit the dimensionless prefactor in the expression (\ref{defnut}) of the turbulent viscosity to $c=0.012$, which is of the order of the typical values retained for simple flow geometries \cite{Pope}. We then plot $\hat{\zeta}$ normalized by its value without waves in figure \ref{fig:scalings}b. This prediction agrees rather well with the experimental data, and for fast Stokes drift the decrease in surface vorticity indeed scales approximately as $1/S$. 

We have therefore understood separately the origin of the two phenomena observed experimentally: the reduction in surface vorticity results from diffusive vorticity spreading associated with turbulent viscosity, while the shift of the vortex center in the direction transverse to wave propagation originates from the nonlinear advection term and the self-induced motion of the vortex. We close this section by stressing the fact that, although the Stokes drift  is restricted to a layer of depth $\lambda/4 \pi$ near the free-surface, in both models it impacts the mean-flow over the entire fluid depth: see the right-hand panel of figure \ref{fig:line}, and the $z$-dependence in expressions (\ref{exprM1}-\ref{exprM2}).

\section{Nonlinear refraction \label{secNLref}}


We now discuss the consequences of the vortex distortion for the refraction of surface waves. As discussed in the outset, waves impinging on a vortex get refracted by an angle proportional to the typical core vorticity. Indeed,
consider a wavefront in panel \ref{fig:snapshots}b impinging on the vortex in panel \ref{fig:snapshots}a: the parts of the wavefront travelling on each side of the vortex core see mean-flow velocities of opposite signs and therefore undergo a differential Doppler shift \cite{Coste99bis}. This leads to a distortion of the wavefront by a distance $\delta x \sim R U_0 /(\lambda f)$ along $x$, over a transverse extension $\delta y \sim R$ along $y$. The corresponding angle $\theta \simeq \delta x / \delta y$ between the refracted wavefront and the $y$-axis is then proportional to the mean-flow vorticity: 
\begin{equation}
\theta \sim \zeta R  /(\lambda f) \, ,
\end{equation}
with $\zeta \sim U_0/R$.
A central result of this study is that surface vorticity is strongly reduced by the presence of intense waves: what are the consequences for wave refraction? Is the refraction angle given by the unperturbed bottom vorticity $\overline{\zeta}_0 \sim U_0 /R$, or by the strongly reduced surface vorticity $\overline{\zeta}$? 

\begin{figure}
\centering{ {\includegraphics[width=8cm]{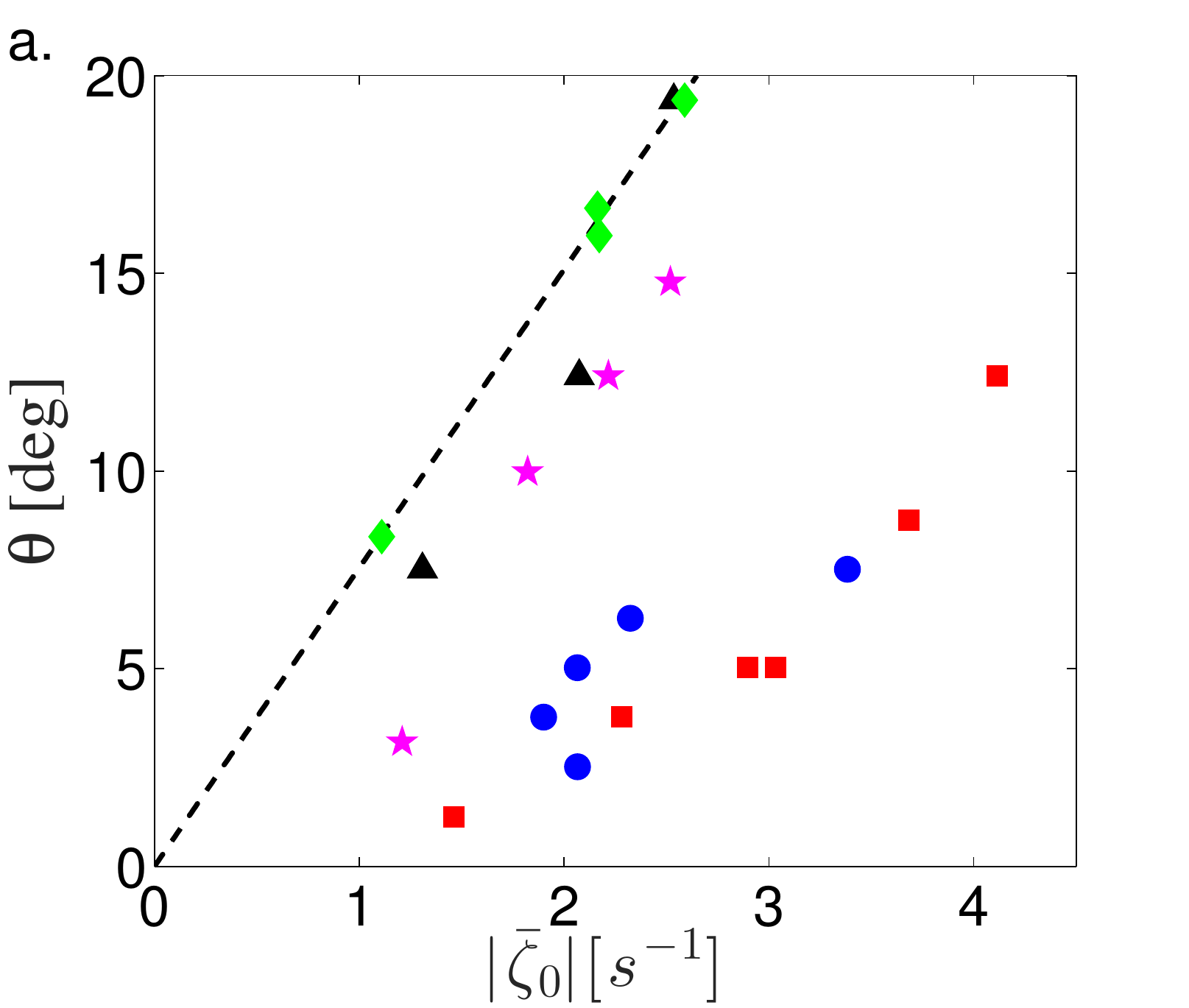}}
     {\includegraphics[width=8cm]{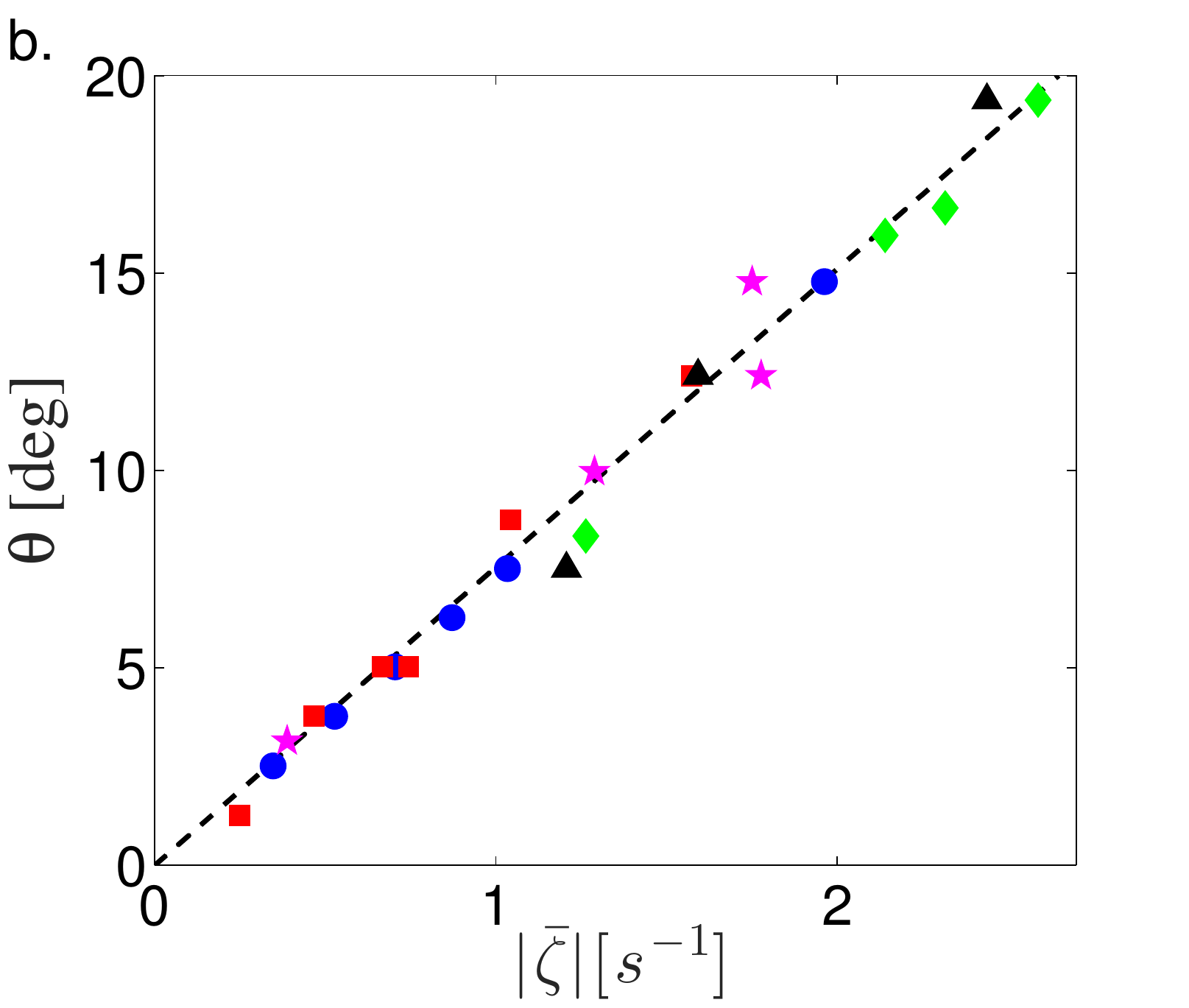}}}
    \caption{Refraction angle $\theta$ as a function of the control core vorticity $\overline{\zeta}_0$ (panel a), and as a function of the core vorticity $\overline{\zeta}$ of the distorted vortex (panel b). Same symbols as in figure \ref{fig:scalings}.
    \label{fig:refraction}}
\end{figure}

Snapshots of the wave field are displayed in figure \ref{fig:snapshots}: while weak waves are strongly refracted by the vortex, with wavefronts departing from the $y$ direction (panel \ref{fig:snapshots}d), for stronger wave amplitude this refraction is reduced (panel \ref{fig:snapshots}f).
To extract quantitative information from these pictures, we define the refraction angle $\theta$ as the average angle between the wavefronts and the $y$-axis, inside the right-most quarter of the images, averaging over all the phases of the wave field. We plot $\theta$ as a function of the base vorticity $\overline{\zeta}_0$ in figure \ref{fig:refraction}a: for weak waves $\theta$ is indeed proportional to $\overline{\zeta}_0$, but as the waves get stronger the refraction angle decreases rapidly. Even more surprisingly, the refraction angle is not linear in $\overline{\zeta}_0$ anymore: the simple law $\theta \sim \overline{\zeta}_0 kR /f$ does not hold for such intense waves.

To solve this puzzle, in figure \ref{fig:refraction}b we plot the refraction angle as a function of the surface vorticity $\overline{\zeta}$ measured in presence of waves: all the data collapse onto a single master curve, with $\theta$ proportional to $\overline{\zeta}$. This shows that the surface-wave field feels only the vorticity located near the free surface. Because strong waves reduce surface vorticity, they are less refracted than weak waves. We conclude that there is a nonlinear regime of wave refraction, where the properties of the refracting medium (the vortex) are strongly affected by the wave field. In terms of control parameters, the law $\theta \sim \overline{\zeta}_0 kR/f$ is valid for waves with $S \ll 1$ only; for intense waves, it must be replaced by $\theta \sim \overline{\zeta} kR/f$. \corr{In the experiment the surface vorticity decreases approximately as $1/S$ for strong waves, a scaling behavior predicted by the simplified expulsion model of section \ref{skin}: as a consequence, the refraction angle decreases as $1/S$ as well, i.e., it is inversely proportional to the wave intensity. It would be interesting to investigate whether this law extends further into the strong-wave regime $S \gg 1$.}


\section{Conclusion}

We have reported the direct experimental observation of wave-induced vortex distortion: the incoming wave field shifts the position of the vortex center and dramatically reduces surface vorticity. We have interpreted these observations in the framework of the Craik-Leibovich equation and concluded that the relevant control parameter is the dimensionless Stokes drift $S$. Using a strongly idealized  vortex-line model, we showed that the shift of the vortex center in the direction transverse to wave propagation can be understood as a balance between Stokes drift and self-induced vortex velocity, which leads to a vortex center displacement proportional to $S$ for small $S$. The decrease in surface vorticity results from vorticity expulsion by the fast Stokes drift near the free surface, a process analogous to the skin effect (i.e. magnetic field expulsion) in an MHD context. \corr{Using a simplified model valid in the limit of fast Sokes drift and shallow fluid layers, we showed that the typical surface vorticity decreases as $1/S$, a prediction compatible with the experimental data.}

These results have important consequences for wave refraction: while weak waves are strongly refracted by vorticity, intense waves expel the background vorticity from the near-surface region and are therefore only weakly refracted. This mechanism leads to a nonlinear regime of wave refraction, where the refraction angle rapidly decreases with wave intensity. 
This nonlinear refraction is probably a general phenomenon arising for other types of waves interacting with vortical flows.

\corr{It would be interesting to investigate whether such wave-induced vortex distortion and nonlinear wave refraction are of relevance to natural flows. An important step in this direction would be to consider free vortices instead of the continuously forced ones of the present experiment.} \\

{\it Acknowledgments.} We thank W.R. Young for many insightful discussions \corr{and for pointing out the possibility to derive the hierarchy of equations (\ref{eqM0}-\ref{eqM2})}, and Vincent Padilla for building part of the experimental setup. We acknowledge support from Labex PALM ANR-10-LABX-0039 and ANR Turbulon 12-BS04-0005.

\appendix

\section{Craik-Leibovich theory for fast mean-flows \label{appendix}}

In the original Craik-Leibovich theory, the waves arise at first order in wave-slope, whereas the mean-flow arises at second order in wave-slope. In this appendix we show that the same Craik-Leibovich equation applies when the mean-flow velocity is of first order in wave slope, i.e., of the same order as the wave orbital velocity.
The starting point of the Craik-Leibovich expansion is the vorticity equation:
\begin{equation}
\partial_t \bomega = \bnabla \times \left( {\bf u} \times \bomega \right) + \nu \bnabla^2 \bomega  \, .\label{NSvort}
\end{equation}
Consider waves having a weak slope, which defines a small expansion parameter $\epsilon$. Expand the velocity and vorticity fields according to:
\begin{eqnarray}
{\bf u} & = & \epsilon {\bf u} _1 + \epsilon^2 {\bf u}_2  + \epsilon^3 {\bf u}_3 + \dots \, ,\\
\bomega & = & \epsilon \bomega _1 + \epsilon^2 \bomega_2  + \epsilon^3 \bomega_3 + \dots \, ,
\end{eqnarray}
and introduce the slow timescales $T_1 = \epsilon t$ and $T_2 = \epsilon^2 t$. Finally, consider small viscosity, written as $\nu= \epsilon \tilde{\nu}$.

At order $\epsilon$, equation (\ref{NSvort}) gives simply:
\begin{equation}
\partial_t \bomega_1 = 0 \, .
\end{equation}
We write the solution to this equation as:
\begin{eqnarray}
{\bf u}_1 & = & \tilde{{\bf u}}({\bf x},t) + {\bf U} ({\bf x},T_1,T_2) \, ,\label{CLu1}\\
\bomega_1 & = & \bOmega({\bf x},T_1,T_2) \, .
\end{eqnarray}
Here $\tilde{{\bf u}}$ denotes an irrotational surface wave field, ${\bf U}$ is a slowly evolving mean-flow, and $\bOmega= \bnabla \times {\bf U}$ is the mean-flow vorticity. As compared to the original analysis of \citet{Craik1976}, we introduce the mean-flow directly in (\ref{CLu1}) and not at order $\epsilon^2$: the mean-flow and the wave orbital velocity have comparable magnitudes.
The order $\epsilon^2$ equation is:
\begin{equation}
\partial_t \bomega_2 + \partial_{T_1} \bomega_1 = \bnabla \times \left( {\bf u}_1 \times \bomega_1 \right) + \tilde{\nu} \bnabla^2 \bomega_1  \, . \label{eqvortO2}
\end{equation}
We denote as $\la \cdot \ra$ the average over the fast time variable $t$. Averaging equation (\ref{eqvortO2}), using $\la \tilde{ \bf u}\ra = 0$, leads to:
\begin{equation}
\partial_{T_1} \bOmega = \bnabla \times \left( {\bf U} \times \bOmega \right) + \tilde{\nu} \bnabla^2 \bOmega  \, , \label{dT1Omega}
\end{equation}
which is the standard vorticity equation for the mean-flow. Subtracting this equation from (\ref{eqvortO2}) gives:
\begin{equation}
\partial_t \bomega_2 = \bnabla \times \left( \tilde{{\bf u}} \times \bOmega \right)   \, ,
\end{equation}
which we integrate into:
\begin{equation}
\bomega_2 = \bnabla \times \left( {\bxi} \times \bOmega \right)   \, ,
\end{equation}
where we defined the wavy displacement $\bxi$ such that $\partial_t \bxi = \tilde{{\bf u}}$.

Finally, averaging over $t$ the order $\epsilon^3$ equation leads to:
\begin{equation}
\partial_{T_2} \bOmega = \bnabla \times \la   \tilde{{\bf u}} \times \left[ \bnabla \times \left( \bxi \times \bOmega  \right)   \right]    \ra  = \bnabla \times ({\bf u}_s \times \bOmega  )  \, , \label{dT2Omega}
\end{equation}
where the last equality involves standard manipulations described in \citet{Craik1976}. Collecting the contributions (\ref{dT1Omega}) and (\ref{dT2Omega}) from the successive slow time variables, we obtain the standard Craik-Leibovich equation (\ref{CLint}). To conclude, if the mean-flow is comparable to the orbital velocity, the vortex-force term is a small correction to the usual vorticity equation, whereas if the mean-flow is $\epsilon$ times slower than the orbital velocity then the vortex-force is of the same order as the advective term of the vorticity equation. In any case, equation (\ref{CLint}) is valid.

\section{\corr{Vorticity spreading in the simplified expulsion model} \label{appmoments}}


For simplicity, we assume that the horizontal extent of the fluid domain is much greater than $R$. The domain can therefore be considered infinite in $x$ and $y$. We compute the moments (\ref{defM0}-\ref{defM2}) at the bottom of the fluid domain using the boundary condition (\ref{CLbottom}):
\begin{eqnarray}
M_0 (-H) & = & \frac{\pi R^2}{4} \, \zeta_0 \, , \\
M_1 (-H) & = & 0 \, , \\
M_2 (-H) & = & \frac{\pi R^4}{32} \, \zeta_0 \, .
\end{eqnarray}
We then integrate equations (\ref{eqM0}-\ref{eqM2}) using these boundary conditions, together with the fact that the first $z$-derivative of these moments vanishes at the free surface. We obtain that $M_0$ is independent of $z$, and the following expressions for $M_1(z)$ and $M_2(z)$:
\begin{eqnarray}
 M_1(z) &  = &  \frac{\pi R^2 U_s \zeta_0}{16 k^2 \nu_t}  \left[ e^{-2kH}-e^{2kz} +2k(z+H)   \right]  \, , \label{exprM1}\\
 M_2(z) &  = &  \frac{\pi R^4 \zeta_0}{32} + \frac{\pi R^2 U_s^2 \zeta_0}{32 \, k^4 \nu_t^2} \left[ k(z+H)(2e^{-2kH}+4kH-3) + \frac{3 e^{-4kH}}{4} \right. \label{exprM2} \\
\nonumber & &   \left.  + \frac{e^{4kz}}{4}  -2e^{-2kH} + e^{2kz} (2-2k(z+H)-e^{-2kH} )    \right] \, . 
\end{eqnarray}
The expressions of $X$, $\delta X$ and $\hat{\zeta}$ can be computed from these moments evaluated at $z=0$.

\end{document}